\documentclass[useAMS,usenatbib]{mn2e}
\usepackage{epsfig}
\usepackage{aas_macros}     % for journal abbreviations in references
\usepackage{amsfonts}

\usepackage[usenames]{color}
\usepackage{url}

\newcommand{\remove}[1]{}

\def\be{\begin{equation}}
\def\ee{\end{equation}}
\def\ba{\begin{eqnarray}}
\def\ea{\end{eqnarray}}

\title[An improved model for the nonlinear velocity power spectrum ]
{An improved model for the nonlinear velocity power spectrum }
\author[Elise~Jennings]
{Elise~Jennings$^{1,2}$\thanks{E-mail: ejennings@kicp.uchicago.edu}\\
$^{1}$ The Kavli Institute for Cosmological Physics, University of Chicago, 5640 South Ellis Avenue, Chicago, IL 60637, U. S.\\
$^{2}$ The Enrico Fermi Institute, University of Chicago, 5640 South Ellis Avenue, Chicago, IL 60637, U. S.\\
}

\begin{document}

\date{\today}

\pagerange{\pageref{firstpage}--\pageref{lastpage}} \pubyear{2012}

\maketitle

\label{firstpage}

\begin{abstract}
The velocity divergence power spectrum is a key ingredient in modelling 
redshift space distortion effects on quasi-linear and nonlinear scales. 
We present an improved model for the $z=0$ velocity divergence auto and cross power spectrum 
which was originally suggested by Jennings et al. 2011.
Using numerical simulations we measure the velocity fields 
using a Delaunay tesselation and obtain an accurate prediction of the velocity divergence power spectrum on scales 
$k < 1 h$Mpc$^{-1}$.  
 We use this 
to update the model which is now accurate to 2\% for both $P_{\theta \theta}$ and $P_{\theta \delta }$ at $z=0$ on scales
$k <0.65 h$Mpc$^{-1}$ and $k <0.35 h$Mpc$^{-1}$ respectively.
We find that the formula for the redshift dependence of the velocity divergence power
spectra proposed by Jennings et al. 2011 recovers the measured $z>0$ $P(k)$ to markedly greater
accuracy with the new model. The nonlinear $P_{\theta \theta}$ and $P_{\theta \delta }$ at $z =1$ are recovered
 accurately to better than 2\% on scales
$k<0.2 h$Mpc$^{-1}$.
Recently it was shown that the velocity field 
shows larger differences between  modified gravity cosmologies and $\Lambda$CDM 
compared to the  matter field.
An accurate model for the velocity divergence  power spectrum, such as the one presented here,
 is a valuable tool for analysing redshift space distortion effects in future galaxy surveys and for constraining deviations from general relativity. 
\end{abstract}


\section{Introduction}

\label{sect:intro}

In the hierarchical model of structure formation, gravitational collapse and the accelerating cosmic expansion are two competing 
effects which determine the rate at which structures, such as galaxies and clusters, 
grow in the Universe. In addition to the Hubble 
flow, galaxies possess peculiar velocities, arising from inhomogeneities in the local density field, which can be used
to probe the growth rate of structure \citep{Percival:2007yw, Guzzo, betal2010,  betal2011,BOSS,beutler2012}.
 These peculiar velocities distort the clustering signal along the line of sight giving rise to redshift space distortions
\citep[see e.g.][]{kaiser1987,hamilton1998}. 
It has been shown that a  key ingredient to improve models of the  power spectrum in redshift space is 
 the inclusion of the nonlinear velocity divergence power spectrum \citep{scoccimarro2004,jbp2011a,jbp2011b}.
In this paper we present an improvement to the model for both the auto and cross velocity divergence power spectrum
presented in \citet{jbp2011a}.  The refinement of the  model is 
driven by using a volume weighted Delaunay tesselation method to measure 
the nonlinear velocity fields to greater accuracy and to smaller scales then presented in \citet{jbp2011a}.

Measurements of the growth rate of structure can be used to determine if the accelerating expansion 
is the result of a dark energy component which behaves as a repulsive form of gravity or if
 Einstein's theory of gravity breaks down on cosmological scales \citep[see e.g. ][]{bz2008}.
Independent measurements of the growth rate can be obtained by measuring the clustering of galaxies in redshift space and 
recently there has been renewed interest in improving the models for the clustering signal in redshift space
 \citep{scoccimarro2004,pw2009,t2010,jbp2011a,seljak2011,tang2011,rw2011,kll2012}.
Also recently it has been pointed out that the velocity divergence power spectrum is a more sensitive probe of 
modified gravity than the nonlinear matter power spectrum \citep{j2012} but models for the 2D redshift space
power spectrum are currently not accurate enough to 
allow this quantity to be extracted to a high precision. However, given an accurate model for 
the velocity divergence power spectrum in a $\Lambda$CDM cosmology, 
this can be used to construct predictions for the moments of the power spectrum in redshift space which can be directly 
compared with data and
 has been shown to be accurate in
detecting deviations from general relativity using simulations \citep{jbp2011b,j2012}.

The linear continuity equation, $\nabla \cdot v = - a H f \delta$, gives 
a one to one correspondence between the velocity and density fields where the linear growth rate $f = {\rm d ln} \delta/{\rm d ln} a$ is 
a cosmology dependent factor, $\delta$ is the matter overdensity, $v$ is the peculiar velocity and $H$ is the Hubble parameter.
Once  overdensities become nonlinear, this relationship no longer holds. \citet{b1992} 
derived the nonlinear relation between $\delta$ and
$\nabla \cdot v$ in the case of an initially Gaussian field.
 Many authors since then have extended this relation using e.g. higher order pertubation theory or the spherical collapse model
\citep[see e.g.][]{sf1996,cl1997,k2000,bc2008,k2011}.
\citet{cc2004} were the first
to apply numerical simulations to model the nonlinear velocity and density
spectra, instead of using perturbation theory. Using a grid-based 
pressureless hydrodynamic code their simulations had limited resolution but were able to fit the ratios up to
 $k=1h$Mpc$^{-1}$ and show that perturbative predictions fail at 
$k >0.2h$Mpc$^{-1}$.

The density velocity relation considered in this paper was first 
presented in \citet{jbp2011a}, where the parameters were obtained by 
fitting  to a mass weighted measurement of the velocity divergence field. 
In this paper we present updated parameters obtained by fitting 
to the volume weighted velocity divergence power spectra which have been measured accurately on 
smaller scales than was possible with the mass weighted method \citep{PS2009}. We also show that the model proposed by 
\citet{jbp2011a} for the redshift dependence of the density velocity relation  recovers
 the velocity power spectra at $z>0$ with greater precision
using the new parameters.

This paper is organised as follows: In Section~\ref{sims} we   describe the
 $N$-body simulations used in this paper. In Section~\ref{measuring_v} we discuss two methods for measuring the 
velocity divergence power spectrum.
In Section \ref{RESULTS} we present the main results of this paper including the updated parameters for the nonlinear velocity divergence power spectra.

\section{Measuring $P_{\theta \theta}$ and $P_{\theta \delta}$ from N-body simulations}

\label{sect:fr}
In section \ref{sims} we present the details of the N-body simulations carried out and 
discuss several methods which can be used to measure the velocity divergence power spectrum in Section \ref{measuring_v}.
We focus on a  Delaunay tesselation approach and show the accuracy which with this method can recover 
$P_{\theta \theta}$ and $P_{\theta \delta}$.

\subsection{Simulation details}

\begin{figure*}
{\epsfxsize=13.5truecm
\epsfbox[86 542 493 715]{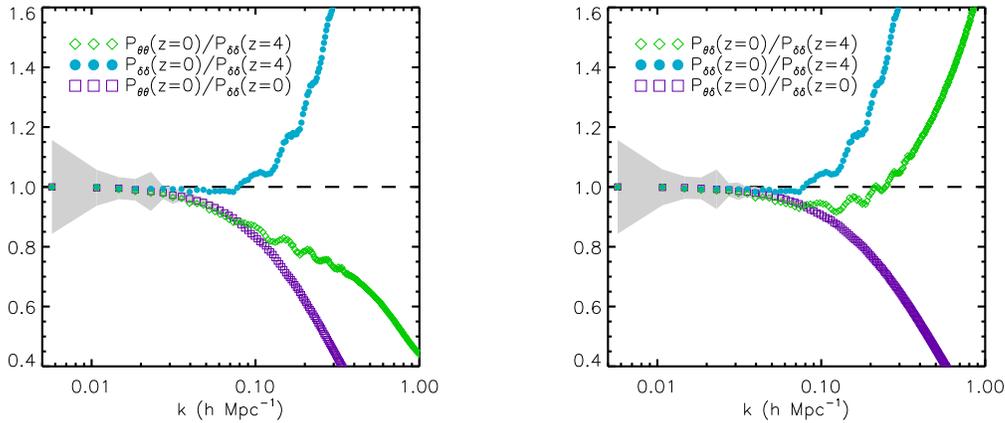}}
\caption{Left panel: Ratios of the $z=0$ nonlinear matter, $P_{\delta \delta}$, and velocity divergence, $P_{\theta \theta}$,
power spectra  to the matter power spectrum at $z=4$, $P_{\delta \delta}(z=4)$, scaled using the ratio of the
square of the linear growth factor at $z = 4$ and $z = 0$ for
$\Lambda$CDM,  measured from six simulations are shown as
filled blue circles and empty green diamonds respectively. The ratio of the $P_{\theta \theta}/P_{\delta \delta}$ measured
from the simulations at $z=0$ is shown as empty purple squares. The shaded grey region shows the errors on this
ratio measured from the scatter amongst six simulations.
Right panel: Similar to the left panel.
The ratios $P_{\theta \delta}(z=0)/P_{\delta \delta}(z=4)$ and  $P_{\theta \delta}(z=0)/P_{\delta \delta}(z=0)$
are shown as empty green diamonds and purple squares respectively.
Note $\theta  = - \nabla \cdot v /(a H f)$.
\label{pttptd}}
\end{figure*}

\label{sims}

 We use the N-body simulations carried out by \citet{lztk2012}.
These simulations were performed at the Institute of Computational Cosmology using 
a  modified version of 
 the
mesh-based $N$-body code {\tt RAMSES} \citep{ramses}.
Assuming  a $\Lambda$CDM cosmology, 
the following cosmological parameters were used in the simulations:
$\Omega_{\rm m} = 0.24$,
 $\Omega_{\rmn{DE}}=0.76$, 
$h = 0.73$ and a spectral tilt of $n_{\mbox{s}} =0.961$ \citep[in agreement with e.g.][]{s2009}.
The  linear theory rms fluctuation
in spheres of radius 8 $h^{-1}$ Mpc is set to be  $\sigma_8 = 0.769$.

The simulations use $N=1024^3$ dark matter  particles to represent the  matter distribution in a  computational box of
comoving length $1500 h^{-1}$Mpc.
The initial
conditions were generated at $z=49$ using the  \citet{mpgrafic} code.
Note the nonlinear matter power spectrum is measured from the simulations  
by assigning the particles to a mesh using the cloud in cell (CIC) assignment scheme  and
performing a fast Fourier transform (FFT) of the density field.
To compensate for the mass assignment scheme we perform an approximate de-convolution following  \citet{bf1991}.

Throughout this paper the velocity divergence is normalized to $\theta  = -\nabla \cdot v /(a H f)$, 
where $v$ is the peculiar velocity, 
$f = {\rm{d ln}} \delta/{\rm d ln} a$ is the linear growth rate, $H$ is the Hubble parameter and $a$ is the scale factor.
Using this normalization $\theta$ is dimensionless and the linear continuity equation is given by $\theta = \delta$.

\subsection{Measuring the velocity divergence field}
\label{measuring_v}

Measuring the velocity divergence field accurately from numerical simulations on small scales 
can be difficult if a mass weighted approach is used as in \citet[][for example]{scoccimarro2004,PS2009, jbp2011a}. 
Some volume weighted measures of the velocity field have also been developed \citep[see e.g.][]{bw1996, cct2007} including the
Delaunay tessellation field estimator (DTFE) method \citep{sv2000, vis2009,cv2011}.

\begin{figure*}
{\epsfxsize=13.5truecm
\epsfbox[88 373 491 541]{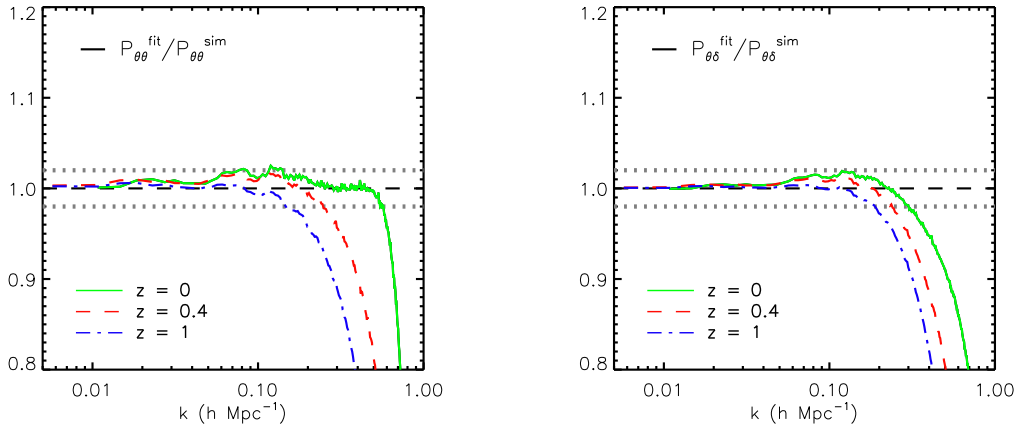}}
\caption{Left panel: The ratio of the fitting formula given in Eq. \ref{g}
for the velocity divergence $P(k)$ to the measured power spectrum,
$P_{\theta \theta}^{{\rm \tiny fit}}/P_{\theta \theta}^{{\rm \tiny sim}}$, at $z=0$ (green solid),
$z=0.4$ (red dashed) and $z=1$ (blue dot dashed).
Right panel: Similar results as shown in the left panel but for
the cross power spectrum $P_{\theta \delta}$.
The horizontal dotted grey lines show a region of 2\% accuracy in the fit.
\label{pz}}
\end{figure*}

In the mass weighted approach, simply interpolating the velocities 
to a grid, as suggested by \citet{scoccimarro2004}, gives the momentum field which is then Fourier 
transformed and divided by the Fourier transform of the density field, which results in a mass weighted velocity field on the grid.
One of the main problems with this approach is that the velocity field is artifically set to zero in regions where 
there are no particles, as the density is zero in these empty cells. 
\citet{PS2009} also found that this method does not accurately recover the 
input velocity divergence power spectrum on scales $k>0.2 h$Mpc$^{-1}$ interpolating the velocities of 640$^3$ particles to a $200^3$ grid.
Using simulations of 1024$^3$ particles in a 1.5$h^{-1}$Gpc box, \citet{jbp2011a}  found that the maximum 
grid size that could be used was 350$^3$ without reaching the limit of empty cells. 

The limit on the maximum grid size  which can be used in a mass weighted estimate of $\theta$ means that the velocity divergence $P(k)$
presented in \citet{jbp2011a} is only accurate on large scales $k<0.2 h$Mpc$^{-1}$. Also, it is not clear on what scales the discrepancy 
between the actual and measured velocity divergence power spectrum, which was found by \citet{PS2009} 
on scales $k<0.02 h$Mpc$^{-1}$ (see  Figure $13$ in that paper), is present in the velocity $P(k)$ measured by \citet{jbp2011a}. 
However, \citet{jbp2011a} find that the measured $P_{\theta \theta}$ and $P_{\theta \delta}$ 
agree with the linear continuity equation on large scales within the errors from eight simulations (see Figure $3$ in that work). 

The limitations of using the mass weighted method to measure $P_{\theta \theta}$ implies that the fitting formula presented in 
\citet{jbp2011a} is only valid over a limited range of scales. The main aim of this paper is to present an improved version of the  formula for $P_{\theta \theta}$
and $P_{\theta \delta}$
given by \citet{jbp2011a} by fitting to  power spectra which have been measured using the 
DTFE method
This code  constructs the Delaunay tessellation from a discrete set of points and interpolates the field values onto a user defined grid.
For the $L_{\tiny \mbox{box}} = 1500h^{-1} $ Mpc simulation
we  generate the velocity auto, $P_{\theta \theta}$, and cross power spectrum, $P_{\delta \theta}$, on a $1024^3$ grid.
The velocity divergence field is interpolated onto the
grid by  randomly sampling the field values at a given number of sample points within the Delaunay cells and
than taking the average of those values.
The resolution of the mesh used in this study means that
mass assignment effects are negligible on the scales of interest here.

In Fig. \ref{pttptd} we show the ratios of $P_{\delta \delta}(z=0)/P_{\delta \delta}(z=4)$ and 
$P_{\theta \theta}(z=0)/P_{\delta \delta}(z=4)$  which represent the average ratio measured from six simulations 
as filled blue circles and empty green diamonds respectively. Here  $P_{\delta \delta}(z=4)$ has been scaled using the ratio of the
square of the linear growth factor at $z = 4$ and $z = 0$ for $\Lambda$CDM. Plotting the ratio in this way removes the sampling variance
\citep{be1994} and shows the agreement of the measured velocity and matter $P(k)$ with the 
predictions of the linear continuity equation on large scales, $k < 0.02h$Mpc$^{-1}$ for 
$P_{\theta \theta}(z=0)/P_{\delta \delta}(z=4)$ and $k < 0.09h$Mpc$^{-1}$ 
for $P_{\delta \delta}(z=0)/P_{\delta \delta}(z=4)$. 
This result agrees with similarly 
measured ratios from simulations for the matter 
\citep[see e.g.][]{a2008} and the velocity divergence power spectrum \citep{jbp2011a,j2012,li2012}.

The ratio of $P_{\theta \theta}(z=0)/P_{\delta \delta}(z=0)$ is plotted in the left panel of 
Fig. \ref{pttptd} as empty purple circles. The grey shaded region shows the errors measured 
from the scatter amongst six simulations. In the right panel of Fig. \ref{pttptd} we plot similar ratios as in the left panel
 but for the velocity divergence cross power spectrum $P_{\theta \delta}$. Note the increase in the ratio of 
$P_{\theta \delta}(z=0)/P_{\delta \delta}(z=4)$ on scales $k >0.1 h$Mpc$^{-1}$ which is due to the nonlinear growth in the matter field 
on these scales.
Overall the results plotted in Fig.
\ref{pttptd} show that the velocity divergence field computed using the DTFE method
agrees with the predictions of linear theory on extremely large scales and allows us to accurately measure the power spectra 
on nonlinear scales provided we use a large enough grid to interpolate the velocities.
We have checked that the effect on the measured velocity power spectra using smaller grid sizes such as $256^3$ and $512^3$
and find excellent agreement between the $P(k)$ measured using a $512^3$ and $1024^3$ grid on scales $k<1 h$Mpc$^{-1}$; the $P(k)$
 measured using a $256^3$ grid shows deviations at $k \sim 0.7 h$Mpc$^{-1}$.

\section{Results \label{RESULTS}}

Using the measured nonlinear velocity divergence auto and cross power spectrum  on a $1024^3$ grid
presented in Section \ref{measuring_v}
and the nonlinear matter power spectrum measured from the simulations at $z=0$, we can fit for new parameters in the following model 
\citep{jbp2011a}:
\begin{eqnarray}
\label{g}
P_{x y}(k) = g(P_{\delta \delta}(k)) = \frac{\alpha_0\sqrt{P_{\delta \delta}(k)}  +\alpha_1 P_{\delta \delta}^2(k)}{\alpha_2 + \alpha_3 P_{\delta \delta}(k)} \, ,
\end{eqnarray}
where $P_{\delta \delta}$ is the nonlinear matter power spectrum. We obtain the following: 
\begin{itemize}
\item $P_{x y} = P_{\delta \theta}$: 

$\alpha_0 = -12483.8$, $\alpha_1 = 2.554$, $\alpha_2 = 1381.29$, $\alpha_3 = 2.540$;

\item $P_{x y} = P_{\theta \theta}$:

$\alpha_0 = -12480.5$, $\alpha_1 = 1.824$, $\alpha_2 = 2165.87$, $\alpha_3 =1.796 $;
\end{itemize}
all points were weighted equally in the fit.
Note the parameters $\alpha_{0-3}$ are not dimensionless and in this work their units differ from those quoted in \citet{jbp2011a}
as the best fit parameters presented here generate
 velocity divergence power spectra which are normalized as $\theta  = \nabla \cdot v /(a H f)$. Note the growth rate at $z=0$ for 
the $\Lambda$CDM cosmology considered in this work is $f = 0.452$. 
We do not give the units of $\alpha_{0-3}$ here but these can be easily found given this normalisation for theta.
The power spectra used for this fit are the average $P_{\theta \theta}$, $P_{\delta \theta}$ and $P_{\delta \delta}$ measured from
six $\Lambda$CDM simulations.

In Fig. \ref{pz} we plot the ratio of $P_{\theta \theta}^{{\rm \tiny fit}}/P_{\theta \theta}^{{\rm \tiny sim}}$ and 
$P_{\theta \delta}^{{\rm \tiny fit}}/P_{\theta \delta}^{{\rm \tiny sim}}$ in the 
left and right panels respectively at $z=0 $ as a green solid line. Here 
the prefix \lq fit\rq\, denotes the velocity divergence power spectrum 
found using the new best fit parameters and the nonlinear matter $P(k)$ measured from the simulations 
 in Eq. \ref{g}. The prefix \lq sim\rq \, denotes the nonlinear velocity divergence measured directly from the simulations using the DTFE.
The formula given in Eq. \ref{g} is accurate to 2\% (shown as the region enclosed by the dotted gray lines in Fig. \ref{pz}) on scales 
$k <0.65 h$Mpc$^{-1}$ for $P_{\theta \theta}$ and $k <0.35 h$Mpc$^{-1}$ for $P_{\theta \delta}$ at $z=0$.
Note the lower limit for the domain of the function given in Eq. \ref{g} is $k \geq 0.006 h$Mpc$^{-1}$.
Although the values for $\alpha_{0-3}$ were obtained by fitting to a $\Lambda$CDM simulation with a particular set of cosmological
parameters, \citet{jbp2011a} found that this relation between the density and velocity power spectra is 
quite insensitive to both the cosmological model, for smooth dark energy models such as quintessence, and the choice of \
cosmological parameters. These results agree with previous studies such as \citet{b1992} and
\citet{b1995}.

The redshift dependence of $P_{\theta \theta}$ and $P_{\theta \delta}$ can be described using the following formula
\citep{jbp2011a}:
\begin{eqnarray}
\label{fullmodel}
P_{x y }(k,z') = \frac{ g(P_{\delta \delta}(k,z=0))  - P_{\delta \delta}(k,z=0)}{ c^2(z=0, z') } \nonumber \\
+ P_{\delta \delta}(k,z') \, ,
\end{eqnarray}
where $g(P_{\delta \delta})$ is the function in Eq. \ref{g}
 and $P_{x y }$ is either the nonlinear cross or auto power spectrum,
$P_{\delta \theta}$ or $P_{\delta \delta}$ and the function $c$ is given by
\begin{eqnarray}
\label{c}
c(z, z') = \frac{D(z) + D^2(z) +D^3(z)}{D(z') + D^2(z')+ D^3(z')} \, ,
\end{eqnarray}
and $D(z)$ is the linear growth factor.

In Fig. \ref{pz} we show the ratio of $P_{\theta \theta}^{{\rm \tiny fit}}$ from 
 Eqns. \ref{g} \& \ref{fullmodel} to the measured velocity divergence power spectrum $P_{\theta \theta}^{{\rm \tiny sim}}$ 
at $z=0.4$ (red dashed) and $z=1$ (blue dot dashed). All of the results shown in this plot represent the average over six $\Lambda$CDM 
simulations. It is clear from Fig. \ref{pz} that the model given in Eq. \ref{fullmodel} is accurate to better than 2\% on scales 
$k<0.2 h$Mpc$^{-1}$ and to 20\% for $ 0.2 < k (h$Mpc$^{-1}$$)< 0.4$ at $z=1$.

\section{Summary \label{CONCLUSIONS}}

Measuring the growth of structure using the anisotropic clustering signal in redshift space is an important tool for discriminating between dynamical dark energy or 
 modified gravity and the standard $\Lambda$CDM cosmological model.
The velocity divergence power spectrum has been shown to be an important ingredient in modelling redshift space distortions \citep{scoccimarro2004,jbp2011a} and
shows larger deviations between modified gravity cosmologies and $\Lambda$CDM  than the differences found in the nonlinear matter power spectrum using numerical
 simulations \citep[see e.g.][]{svh2009,j2012,li2012}. 

In this paper we present an improved model for the $z=0$ velocity divergence auto and cross power spectrum which was 
originally suggested by \citet{jbp2011a}. 
By measuring the velocity fields using a Delaunay tesselation we obtain an accurate prediction of the velocity $P(k)$ to $k \sim 1 h$Mpc$^{-1}$ which we use to update the 
parameters for the fitting formula given in Eq. \ref{g}.
 We find that this model is accurate to 2\% on scales
$k <0.65 h$Mpc$^{-1}$ for $P_{\theta \theta}$ and $k <0.35 h$Mpc$^{-1}$ for $P_{\theta \delta}$ at $z=0$.
We find that the formula for the redshift dependence of the velocity power 
spectra given in \citet{jbp2011a} recovers the $z>0$ $P(k)$ to markedly greater 
accuracy with the new parameters. This model for the redshift evolution of $P_{\theta \theta}$ and $P_{\theta \delta }$, given in Eq. \ref{fullmodel},
 is accurate to less than 2\% on scales
$k<0.2 h$Mpc$^{-1}$ and to 20\% for $ 0.2 < k (h$Mpc$^{-1}$$)< 0.4$ at $z=1$.

The improved model presented here accurately describes the nonlinear density velocity relation and allows 
the velocity divergence power spectra to be easily and accurately predicted over a range of scales $k <0.9 h$Mpc$^{-1}$ and redshifts $z<=1$.  
This model will be useful for analysing meaurements of peculiar velocities at high and low redshifts  \citep[e.g.][]{ht2012} 
and for studying the redshift space clustering
 signal  in galaxy redshift surveys. This improved model has already been implemented in 
the analysis of the 6dF galaxy redshift survey \citep{beutler2012}.

\section*{Acknowledgments}

We would like to thank Baojiu Li for allowing us to use the simulations presented in this study 
and Carlton Baugh and Martin Crocce for useful 
comments and discussions.
EJ acknowledges the support of a grant from the Simons Foundation, award number 184549. This work was supported in part by the Kavli Institute for Cosmological Physics at the University of Chicago through grants NSF PHY-0114422 and NSF PHY-0551142 and an endowment from the Kavli Foundation and its founder Fred Kavli.
The calculations for this paper were performed on the ICC Cosmology Machine, which is part of 
the DiRAC Facility jointly funded by STFC, the Large Facilities Capital Fund of BIS, and Durham University.

%\bibliographystyle{mn2e}
%\bibliography{mybibliography}

\bsp

\label{lastpage}

\end{document}